\documentstyle[12pt]{article}

\begin{document}
\pagestyle{empty}
\begin{center}
{\Large\bf Relativistic Corrections to the Triton Binding Energy}
\\
F. Sammarruca,
 D. P. Xu$^\dagger$,
 and
 R. Machleidt\\
 {\it Department of Physics,
  University of Idaho,\\
   Moscow, Idaho 83843, U.S.A.}
\\
{\sc \today}
\end{center}

\begin{abstract}
The influence of relativity on the triton binding energy is investigated.
The relativistic three-dimensional version of the Bethe-Salpeter
equation proposed by Blankenbecler and Sugar (BbS) is used.
 Relativistic (non-separable) one-boson-exchange potentials
(constructed in the BbS framework) are employed for the two-nucleon
interaction.
In a 34-channel Faddeev calculation, it is found that relativistic
effects increase the triton binding energy by about 0.2 MeV.
Including charge-dependence (besides relativity), the final triton binding
energy predictions are 8.33 and 8.16 MeV for the Bonn A and B
potential,
respectively.
\end{abstract}

{\footnotesize
$^\dagger$ Present address: Department of Physics, Massachusetts Institute
of Technology, Cambridge, Massachusetts 02139.}

\pagebreak

\setcounter{page}{1}
\pagestyle{plain}
\section{Introduction}

One of the main objectives of
nuclear few-body physics is to learn something
about nuclear forces. When the few-body problem can be solved accurately,
results can be related directly to the
elementary (two- and many-body) forces used
as input in the calculations.
Since the two-nucleon scattering data determine the
nucleon-nucleon (NN) interaction
only on-shell, one has to consider A-nucleon systems with $A>2$
to test off-shell properties of nuclear potentials.
Similarly, many-body forces can only be tested in systems with $A>2$.
Guided by these ideas,
comprehensive work has been performed for, particularly, the
trinucleon ($3N$) system.
For this system, an exact and practically feasible theory has been
developed by Faddeev~\cite{Fad61}.

In recent years, many benchmark calculations of the triton binding energy
employing realistic $NN$ potentials have been performed~\cite{2BF,WGK91,Bra88}
 --- with
the objective of obtaining (indirect) information on the off-shell
behaviour of the two-nucleon potential and to assess the size of the
nuclear three-body force. It has been shown that
when two-nucleon forces only are applied, the triton is underbound
by an amount which varies from 0.2 to 1.1 MeV (depending on which NN
potential is used in the calculation).
 The conventional interpretation of this result
is that it `proves
empirically' the existence of three-body forces
(which would contribute  $ 0.2$ MeV or more to the triton binding). As shown
in many calculations~\cite{3BF}, it is then possible
to patch up the discrepancy with experiment by including
phenomenological
 three-body forces which can easily be fit to any amount of missing binding
energy. Unfortunately,
not much physics can be learned in this {\it ad hoc} fashion.

Strictly speaking, most many-body forces are an artefact of theory. They
are created by freezing out degrees of freedom contained in the full
Hamiltonian of the problem under consideration.
This fact suggests how one may learn some physics
from three-body-force calculations:
when a new degree of freedom is introduced (besides the nucleon),
it should be taken into account consistently in the two- {\it and}
the three-body problem. The theory of the NN
interaction has shown that isobar degrees of freedom
(particularly the $\Delta(1232)$) are crucial
for a realistic and reliable description of the nuclear force~\cite{MHE87}.
Trinucleon calculations taking the
$\Delta$-degree of freedom consistently into account
have been performed first by the Hannover group~\cite{HSS83}.
Recently, this type of calculation has been updated and improved~\cite{PRB91}.
The new result is that the sum of all three-body force contributions
arising from
the $\Delta$-isobar is repulsive; the repulsive
contribution to the triton binding energy varies between 0.06 and 0.4 MeV
depending on the $\Delta$ model~\cite{PRB91}.
This small total result is due to a cancelation
between an attractive three-body force contribution of about 1.5 MeV and
a slightly larger repulsive dispersive effect on the $\Delta$-diagrams
contained in the two-nucleon interaction.
Even if the $\pi-N$ $S$-wave part
of a $2\pi$-exchange three-body force~\cite{FGPC88} (not included in
 a $\Delta$ model)
is added, the total result is essentially vanishing.
Thus, there is little hope that a {\it consistent} theory
for two- and three-body forces will explain the triton
binding energy within the conventional framework.

On the background of the above discussion,
it becomes increasingly important to investigate aspects
left out in conventional three-nucleon
calculations.
One of these aspects is relativity.
Most present
calculations using realistic two-body forces are performed
in a non-relativistic framework. Thus, one may ask whether relativistic
effects are responsible for the missing binding.
Since the triton wave functions contain high momentum components, it is
by no means clear that the non-relativistic approximation is justified.
Furthermore, some non-relativistic predictions
for the
triton binding energy
(obtained with two-body forces only)
are only 0.2 MeV off the empirical value~\cite{Bra88}.
Thus, even if relativistic effects are small, they may be large as
compared to the small gap to be filled by `non-conventional' nuclear
physics.
 In summary, the current status of the triton binding energy program
calls for an accurate knowledge of the relativistic corrections.

In the past 25 years, there have been numerous efforts to estimate
relativistic effects in the three-nucleon system.
Roughly, one can distinguish between two lines of research.
In a model independent approach, one starts from the Poincar\'{e}
algebra for the ten generators of the Poincar\'{e}
group, expressed in terms of the degrees of freedom of $n$ particles
only. This frame work is applied in Refs.[10-13] and critically examined in
Refs.[14-16].

An alternative approach is to use a field-theoretic framework and
the Bethe-Salpeter
 equation~\cite{SB51} or one of its three-dimensional reductions.
Pioneering work along this line has been performed by Tjon and
coworkers~\cite{AT66,JT70}.
In reference~\cite{JT70},
a Blankenbecler-Sugar~\cite{BS66} three-body equation for a modified
Reid potential is solved. In a series of recent studies, Rupp and Tjon
solved the full four-dimensional three-body Bethe-Salpeter equation
{}~\cite{RT88,RT92}. However, to make this tremendous numerical project
feasible,
finite-rank separable NN potentials had to be used.
In these studies, relativistic effects are found to
increase the triton binding by
about 0.3 MeV.

In the present paper,
we investigate the influence of relativity in the $3N$ bound-
state system using the relativistic meson-theoretic Bonn
potential~\cite{MHE87,Mac89}.
It is important that relativistic few-body calculations are performed
consistently. By this we mean that the same relativistic formalism has to be
used for the two- and the three-body systems. Since the relativistic potential
we apply is constructed in the framework of the Blankenbecler-Sugar
equation~\cite{BS66}
(a relativistic three-dimensional reduction of the Bethe-Salpeter
equation~\cite{SB51}),
 we use an equivalent relativistic equation for the three-nucleon system.

 Various three-dimensional
 relativistic three-body equations have been
proposed, which all have in common the elimination of the relative time
component in the original four-dimensional integration over internal momenta.
This is usually done by replacing, in the  three-particle
Bethe-Salpeter equation, the original Green's function by one that
involves a delta function. This procedure is carried out in such a way as
to preserve relativistic invariance, three-particle unitarity and, of course,
the correct non-relativistic limit.

As mentioned above,
our study is performed within such relativistic
 three-dimensional approaches, combining, however,
a non-separable relativistic two-body potential
(the Bonn potential), with relativistic
Faddeev equations.
Among our objectives
it is to understand to which extent, if at
all, the type of two-body potential used is related to the
size of the relativistic
effect.
 Finally, our main motivation is that
 the influence of relativity on the $3N$ bound-state
 properties is a basic and fundamental problem, and we must achieve a
solid understanding of it.

In the next Section,
we will briefly recall the relevant equations and describe the
relativistic corrections. In Section 3 we present and discuss our results.

\section{Formalism}

The relativistic two-body problem is described by the Bethe-Salpeter (BS)
\cite{SB51} equation, which can be extended to
the three-body case. Similarly to
 the non-relativistic formalism,
 the relativistic three-body BS equation can be cast into
Faddeev form~\cite{BS66,AO65,FLN66,AAY68}. Unlike in the
non-relativistic case, however,
 one obtains a four-dimensional integral
equation, (the additional integration variable is due to
the intermediate particles
being off their mass shell),
 which drastically complicates the numerical solution.
 To avoid this difficulty, various three-dimensional reductions have
been worked out as a practical generalization of the relativistic Faddeev
equations. Here, we just briefly recall the procedure.
The relativistic Faddeev equations are written as
\begin{equation}
T^{i}= t_{i} + \sum_{j \neq i} t_{i}G_{0i}T^{j}
\end{equation}
with $i,j=1,2,3$.
 $t_{i}$ are two-body scattering operators describing all
interactions between particles $j$ and $k$, and $G_{0i}$
is the free two-particle
Green's function of the pair $j,k$ acting in the three-body space. (In the
non-relativistic case, it can be shown that $G_{01}=G_{02}=G_{03}$, all
particles being on their mass shell.)
The relativistic propagator, e.g. for particles $j$ and $k$, is then
\begin{equation}
G_{0i}= \frac{1}{(k_{j}^{2}-m^{2}+i\epsilon)} \frac{1}{(k_{k}^{2}-m^{2}
+i\epsilon)}
\end{equation}
with $k_{i}$ the four-momentum of particle $i$ in the three-body
center-of-mass (C.M.) frame.
The usual procedure is then to replace the relativistic four-dimensional
 propagators by a three-dimensional one which is derived from
 a dispersion integral~\cite{AT66,JT70}.
This amounts to introducing a Green's function which produces the
proper right-hand (unitarity) cut.
The procedure
is, however, not unique, and many different versions can be found in the
literature.

Instead of the Green's function Eq.\ (2), one may write a dispersion
integral in terms of the two-particle invariant energy,
$\sigma_{i}$, (as we choose to do in
this work), namely
\begin{equation}
g_{i}=4 \int_{4m^{2}}^{\infty} \frac{d\sigma'_{i}}{\sigma'_{i}-\sigma_{i}}
\delta [(P'_{jk}/2+k_{jk})^{2}-m^{2}]\times
\delta [(P'_{jk}/2-k_{jk})^{2}-m^{2}]
\end{equation}
where $k_{jk}\equiv (k_{j}-k_{k})/2$, $ \sigma _{i}=
(P-k_{i})^{2}=P_{jk}^{2}$
with $P = k_i+k_j+k_k$ the total four-momentum,
 $P_{jk}\equiv k_{j}+k_{k}$,
 and $P'_{jk}=
\sqrt{ \sigma '_{i}/\sigma_{i}}P_{jk}$.
In the three-body C.M. frame, $P=(\sqrt{s},{\bf 0})$ with
$\sqrt{s}=3m-E_b$ the invariant mass of the three-body system ($m$
denotes the mass of a free nucleon
and $E_b$ the binding energy of the three-nucleon
system).
Clearly, this choice cannot be unique, since any function which does not
contribute to the right-hand cut of $\sigma$ can be added
without affecting unitarity.
Evaluation of the integral in Eq.\ (3) results in the three-dimensional
propagator~\cite{AT66,JT70}
\begin{equation}
g= \frac{4}{\omega _{p}} \frac{1}{4\omega _{p}^{2}+|{\bf q}|^{2}-
(\sqrt{s} - \omega_{q})^{2}}
\end{equation}
Here, ${\bf q}({\bf q'})$ are the momenta of the {\it spectator} particle
in the initial(final) state, respectively,
$\omega_{p}=\sqrt{m^{2}+|{\bf p}|^{2}}, {\bf p}=-\frac{1}{2}{\bf q}-{\bf q'},$
and $\omega_{q}=\sqrt{m^{2}+|{\bf q}|^{2}}$.
Notice that this is just a more convenient notation which allows to drop
particles indices, namely ${\bf q}={\bf k}_i,
{\bf q'}={\bf k'}_k$, where $i$ and $k$ are understood to be the
{\it spectator}
particles in the initial and final state respectively.
Clearly, ${\bf q}$ and ${\bf q'}$ are also the relative momenta between the
spectator
and the interacting pair in the three-body C.M., see Fig.\ 1.

Alternatively, one could have written
the dispersion integral in terms of the three-particle invariant
energy, that is
\begin{equation}
g_{i}=4\int_{9m^{2}}^{\infty}\frac{ds'}{s'-s}\delta\{[(P'-k_{i})/2+k_{jk}]^{2}
-m^{2}\}\times \delta\{[(P'-k_{i})/2-k_{jk}]^{2}-m^{2}\}
\end{equation}
with $k_{jk}$ as above, $s=P^{2}$, $P'=P \sqrt{s'/s}$. The resulting
three-dimensional
propagator would then be~\cite{AT66,RT88}
\begin{equation}
g= \frac{2}{\omega _{1} \omega _{2}} \frac{\omega _{1} + \omega _{2}
+ \omega _{3}}{(\omega _{1} + \omega _{2} +\omega _{3})^{2}-s}
\end{equation}
with $\omega _{1}=\sqrt{m^{2}+|{\bf q'}|^{2}}, \omega_{2}=\sqrt{m^{2}+|{\bf q}+
{\bf q'}|^{2}}$ and $\omega_{3}=\sqrt{m^{2}+|{\bf q}|^{2}}$.

Other alternatives have also been suggested. One could write as well
 the dispersion
integral in terms of the one-particle invariant energy, or introduce
relativistic kinematics in the free resolvent of the non-relativistic Faddeev
equation; in the latter case, phase space factors are introduced to make the
volume element of the integration relativistically invariant~\cite{AT66}.
All choices are equivalent in the sense that they satisfy three-body
unitarity and preserve the correct non-relativistic limit.
This non-relativistic Faddeev propagator reads in our notation
\begin{equation}
g_{NR}=\frac{1}{m} \frac{1}{|{\bf p}|^{2}+\frac{3}{4}|{\bf q}|^{2}+mE_{b}}
\end{equation}

Besides the modifications in the propagator, additional corrections must be
included to properly define the dynamical variables. Clearly, in the three-
body calculation, these are defined in the overall C.M. system and must appear
correctly in the input two-body t-matrix. The diagram in Fig.\ 1 is a typical
graphical representation of the three-particle reaction $i(jk) \rightarrow
k(ij)$. The double lines represent interacting pairs (with momenta $-{\bf q},
-{\bf q'}$, respectively).
The total invariant mass of the two-body sub-system is
\begin{equation}
\sigma= (\sqrt{s} - E_{q})^{2} -|{\bf q}|^{2}
\end{equation}
with $E_{q}=\sqrt{m^{2}+|{\bf q}|^{2}}$, so that
$(\sqrt{s} - E_{q})$ is nothing but the energy of the pair with total
three-momentum $-{\bf q}$ (righthandside of diagram in Fig.\ 1).

Also,
 the relative momenta of the interacting pair must be defined in a
relativistic way (in a non-relativistic theory, they would be the Galilean
invariant relative momenta). It has been shown
shown~\cite{AAY68,RT77} that the relativistic
relative momentum, (which reduces to the momentum of one of the particles in
their C.M. system), is given by the combination
\begin{equation}
{\bf Q}= {\bf q} + \rho (|{\bf q}|,|{\bf q'}|,z) {\bf q'}
\end{equation}
(for the lower vertex in Fig.\ 1). In the last expression,
 $z$ is the angle between ${\bf q}$ and ${\bf q'}$, and the function $\rho$
 is given by~\cite{AAY68,RT77}
\begin{equation}
\rho (|{\bf q}|,|{\bf q'}|,z)= \xi _{q'}^{-1/2}
(E_{q}+ \frac{{\bf q}\cdot {\bf q'}}{\xi _{q'}^{1/2} + E_{q} + E_{{\bf q +
q'}}})
\end{equation}
with $ \xi _{q'}=(E_{q} + E_{{\bf q+ q'}}) - {\bf q'}^{2}$.
(A completely analogous definition applies
to the upper vertex, but with ${\bf q}, {\bf q'}$
interchanged.)
Notice that ${\bf Q}, {\bf Q'}$ are formally identical to the Galilean
invariant relative momenta, if the function $\rho$ is replaced by a factor
1/2.

Finally, we mention that the relativistic two-body t-matrix,
$t_{rel}$, is related to the non-relativistic one, $t_{NR}$, by
\begin{equation}
t_{rel}= t_{NR} \sqrt{E/m} \sqrt{E'/m}
\end{equation}
 with $ E=
 \sqrt{m^{2}+{\bf Q}^{2}}$ and $ E'= \sqrt{m^{2}+{\bf Q'}^{2}}$.
If in the relativistic Blankenbecler-Sugar (BbS) equation, $t_{rel}$ is
replaced by Eq. (11) and the quasi-potential, $V_{rel}$, is replaced by
\begin{equation}
  V_{rel}=V_{NR} \sqrt{E/m} \sqrt{E'/m}
\end{equation}
then the usual non-relativistic
 Lippmann-Schwinger equation is obtained~\cite{detail}.

In summary, the relativistic calculation of the triton binding energy
differs from the non-relativistic one in three points:

\begin{enumerate}
\item The relativistic Faddeev propagator, Eq.~(4), is used instead of
the conventional non-relativistic propagator, Eq.~(7).

\item The kinematical variables of the two interacting nucleons are defined
in a covariant way (cf.~Eqs.~(8) and (9)).

\item An invariant two-body t-matrix is used, cf. Eq.~(11).

\end{enumerate}
It will turn out that the last point (invariance of the t-matrix) has
the largest quantitative effect and is responsible for the increase of
the triton binding energy.

\section{Results and Discussion}

In Table 1 we present our results for the non-relativistic and relativistic
calculations of the triton binding energy taking
5 and 34 three-body channels~\cite{foot3} into account.
We apply
the Bonn A and the Bonn B potential~\cite{foot1}
 for
the two-body interaction.
 The relativistic  calculations are based upon the formalism
outlined
in the previous section, with the three-nucleon
propagator given in Eq.~(4).
For the solution of the non-relativistic as well as relativistic
Faddeev equation, we use standard momentum-space techniques~\cite{Wglo77}.

We find a relativistic correction to the triton binding energy of
0.19 MeV for the Bonn A and 0.20 MeV for the Bonn B potential.
As indicated above, the largest effect comes from the use of
an invariant t-matrix; this increases the binding energy by 0.26 MeV.
The effect of using covariant kinematical variables is very small
and further increases the binding energy by 0.07 MeV, while the
relativistic Faddeev propagator, Eq. (4), reduces the binding energy
by 0.14 MeV. (The numbers given here for the partial effects refer
to the 34-channel calculation with Bonn A.)

 Using the Bonn A potential, the non-relativistic
34-channel result of 8.32 MeV is increased to 8.51 by relativity. These numbers
refer to the
charge-independent calculation in which the neutron-proton potential is used
for all states. If the well-established charge-dependence in the $ ^{1}S_{0}$
two-nucleon potential is taken into account, the triton binding energy is
reduced by 0.18 MeV, yielding  the final result of
8.33 MeV for the Bonn A potential.
For Bonn B, the final result, including relativistic effects and
charge-dependence in the $^1S_0$ state, is 8.16 MeV (cf.\ Table~1).
In our calculations, charge dependence is taken into account as described
in Ref.~\cite{Bra88b}.

The relativistic effect we find
is small but not negligible.
Another subtle effect that is usually neglected in triton binding energy
calculations is charge-dependence. Accidentally this effect is of
the same magnitude as the relativistic effect. For this reason, we also
consider charge-dependence in our calculations. Thus, we can provide
rather `complete' final results for the triton binding energy
predictions from the two-body force only. This final result is
that
the triton is underbound by 0.15--0.30 MeV
depending on the choice of the relativistic two-nucleon potential.
This means that these potentials almost completely explain the
triton binding energy in terms of just the two-nucleon force.

At this point, one may raise the question whether it is reasonable
to obtain almost all the triton binding energy from the two-body force alone.
The crucial issue here is what to expect from
three-body forces. As mentioned in the Introduction, the recent
investigations by Picklesimer {\it et al.}~\cite{PRB91}
have confirmed the earlier findings by
the Hannover group~\cite{HSS83} that  a consistent treatment of two- and
three-body
forces results in an almost vanishing total contribution to the triton binding
from three-body forces.
In this light, the triton binding can only be understood
as resulting essentially from the two-nucleon interaction
(unless one wants
to invoke subhadronic degrees of freedom).

Thus, crucial for the understanding of our result is the nature
of the two-body force. It is well-known~\cite{Mac89}
that a weaker tensor force component in the two-nucleon force
implies more attraction in few- and many-body systems.
This is why Bonn A predicts more binding than Bonn B; and, even more so, this
is the reason why both Bonn A and B give substantially
more binding energy than other
conventional potentials which, in general, predict only about 7.5 MeV.
On the other hand,
the tensor force should only be as weak as allowed by the requirement
that the potential be `realistic'.
The usual interpretation of the term `realistic' is that the potential
should reproduce the two-nucleon scattering data up to pion-production
threshold.
This is, in fact,
true for both Bonn A and B, with the exception of the $\epsilon_{1}$ mixing
parameter at energies larger than 100 MeV, which is insufficiently reproduced
by Bonn A, but correctly predicted by Bonn B~\cite{foot2}.
On those grounds, one may be tempted to cast some doubt on the Bonn A
three-body result.

To properly discuss this issue, let us first focus on the Bonn B
predictions. The Bonn B potential is with no doubt realistic (in its
predictions of all two-body bound state and scattering observables,
including the $\epsilon_1$), and
nevertheless predicts substantially more triton binding than other
realistic potentials.
For example, the Paris and the Bonn B potential
yield almost identical predictions for
phase shifts, $\epsilon$ parameters, and observables~\cite{foot4},
i.\ e.\ they are as identical on-shell
 as two different potentials can be.
 However, the triton binding energy prediction
by Bonn B is about 0.4 MeV larger than the one by Paris
(the charge-dependent Paris result is 7.59 MeV~\cite{WGK91}).
This clearly reveals off-shell differences between the two interactions.
The Paris potential is, apart from a $\bf p^2$ operator, a local potential,
while Bonn A and B are defined in terms of the relativistic momentum-space
Feynman amplitudes for one-boson-exchanges (which are
clearly non-local expressions).
These off-shell differences can already be seen in the D-state probability
of the deuteron which is 5.8\% for Paris and 5.0\% for Bonn B.

Concerning now the Bonn  A potential, one has to understand the relevance of
the $\epsilon_1$ parameter for nuclear ground state predictions.
In Reference~\cite{Mac92}, a family of potentials has been
considered that has the same low D-state probability of the
deuteron as Bonn A, but larger $\epsilon_1$ parameter at intermediate energies.
It is found that the triton binding energy is always the same and, thus,
is not affected by the behaviour of
 the $\epsilon_1$ above 100 MeV.

Finally, some comments are in place
concerning the relativistic framework we use.
Given the fact
that our relativistic one-boson-exchange two-nucleon
potentials are constructed within
the BbS scheme, the relativistic three-body equation derived
within the BbS framework
certainly is the most consistent choice.

On the other hand, for the relativistic three-nucleon problem, (unlike the
non-relativistic case),
predictions from state-of-the-art calculations do not corroborate and are, in
fact,
quite method-dependent. For instance,
Rupp and Tjon have shown~\cite{RT88,RT92} that the BbS approximation
yields
about 75\% of the relativistic effect as obtained from
the four-dimensional BS equation.
The reason for this difference may be
the effective three-body forces due to
retardation which are automatically included
in the four-dimensional BS equation
and do not exist in static
approaches.~\cite{RT88,RT92}

Na\"{i}vely, one would believe that
the BS equation is {\it a priori} the `best' equation.
 Based upon this believe, the quality of a relativistic
three-dimensional equation would then be determined by how close it
reproduces results obtained from the BS equation.

However, as pointed out by Gross~\cite{Gro82}, there are {\it physical}
reasons why this seemingly most plausible approach may not be correct.
For example, the BS equation in ladder approximation (that is, with
a kernel which is restricted to one-boson-exchanges) does
not generate the desired one-body equation in the one-body limit
(i.~e. when one of the particles becomes very massive),
while a large family of three-dimensional relativistic reductions of
the BS equation does have the correct one-body limit.
If the sum of all crossed ladders is added to the ladders, then the correct
limit is obtained in the BS equation. As a matter of fact,
in almost all applications in nuclear physics (including the present work)
the kernel is restricted to one-boson-exchanges.
Furthermore, when going beyond the second order kernel,
there is no rapidly converging series of irreducible kernels
in the BS framework~\cite{Gro82}. Better convergence is obtained for
some three-dimensional equations.

Thus, the BS equation may not be the optimal choice for
relativistic investigations in nuclear physics.
Given the problems intrinsic to relativistic approaches, three-dimensional
relativistic equations and results obtained from them may be
equally meaningful or even superior as compared to the
four-dimensional BS theory.

The authors are indebted to R. A. Brandenburg for his momentum-space
Faddeev code.
This work was supported in part by the NSF-Idaho EPSCoR program under
Grant No.~RII-8902065, the NSF Physics program under Grant No.~PHY-8911040, and
by the
San Diego Supercomputer Center.

\newpage

\begin{table}
\caption{Triton binding energy (in units of MeV)
obtained in non-relativistic and relativistic
calculations using two versions of the relativistic Bonn
two-nucleon potential.}
\begin{tabular}{c|c|c|c|c}
\hline\hline
          & Number   &                  &             & Relativistic\\
Potential &  of      & Non-Relativistic & Relativistic&   and       \\
          & Channels &                  &             & Charge-Dependent\\
\hline\hline
          &    5     &   8.37           &   8.53      &  8.35  \\
Bonn A    &          &                 &               &  \\
          &   34     &   8.32           &   8.51      &   8.33 \\
\hline
          &    5     &   8.16           &   8.33      &   8.15 \\
Bonn B    &          &                  &             &   \\
          &   34     &   8.14           &   8.34      &   8.16 \\
\hline\hline
\end{tabular}
\end{table}

\newpage

\begin{figure}[t]
\vspace{5in}
\caption{Graphical representation of the three-body rearrangement collision
 $i(jk) \rightarrow k(ij)$. Double lines represent interacting pairs.
The underlying time axis is horizontal, pointing left into the future.}
\end{figure}


\begin{thebibliography}{99}

\bibitem{Fad61} L. D. Faddeev,
 {\it Soviet Physics.} JETP {\bf 12}, 1041 (1961).
\bibitem{2BF}
E. P. Harper, Y. E. Kim, and A. Tubis, Phys. Rev. C {\bf 6}, 126 (1972);\\
A. Laverne and C. Gignoux, Nucl. Phys. {\bf A203}, 597 (1973);\\
R. A. Brandenburg, Y. E. Kim, and A. Tubis, Phys. Rev. C {\bf 12}, 1368
(1975);\\
R. A. Brandenburg, P. U. Sauer, and R. Machleidt, Z. Physik {\bf A280},
93 (1977);\\
Ch. Hajduk and P. U. Sauer, Nucl. Phys. {\bf A369}, 321 (1981);\\
C. R. Chen, G. L. Payne, J. L. Friar, and B. F. Gibson, Phys.
Rev. C {\bf 31}, 2266 (1985);\\
T. Sasakawa and S. Ishikawa, Few-Body Systems {\bf 1}, 3  (1986).
\bibitem{WGK91}
H. Witala, W. Gl\"{o}ckle, and H. Kamada, {\it Phys. Rev. C} {\bf 43}, 1619
(1991).
\bibitem{Bra88} R. A. Brandenburg, G. S. Chulick, R. Machleidt, A. Picklesimer,
 and R. M. Thaler,
 {\it Phys. Rev. C} {\bf 37}, 1245 (1988).
\bibitem{3BF} A. B\"{o}melberg, Phys. Rev. C {\bf 34}, 14 (1986);\\
C. R. Chen, G. L. Payne, J. L. Friar, and B. F. Gibson, Phys.
Rev. C {\bf 33}, 1740 (1986);\\
S. Ishikawa and T Sasakawa, Few-Body Systems {\bf 1}, 143 (1986).
\bibitem{MHE87} R. Machleidt, K. Holinde, and Ch. Elster,
{\it Phys. Reports} {\bf
149}, 1 (1987).
\bibitem{HSS83} Ch. Hajduk, P. U. Sauer,
and W. Strueve, Nucl. Phys. {\bf A405},
581 (1983);
M. T. Pe\~{n}a, H. Henning, and P. U. Sauer, Phys. Rev. C {\bf 42}, 885 (1990).
\bibitem{PRB91} A. Picklesimer, R. A. Rice, and R. Brandenburg, Phys. Rev.
C {\bf 44}, 1359 (1991); {\it ibid.} {\bf 45}, 547 (1992); {\it ibid.}
{\bf 45}, 2045 (1992).
\bibitem{FGPC88} J. L. Friar, B. F. Gibson, G. L. Payne, and S. A. Coon,
Few-Body Systems {\bf 5}, 13 (1988).
\bibitem{Dir49} P. A. Dirac, {\it Rev. Mod. Phys.} {\bf 21}, 392 (1949);
B. Bamhanjian and L. H. Thomas, {\it Phys. Rev.} {\bf 92}, 1300 (1953);
S. N. Sokolov, {\it Docl. Akad. Nauk USSR}, {\bf 233}, 575 (1977);
F. Coester and W. N. Polyzou, {\it Phys. Rev. D} {\bf 26}, 1348 (1982).
\bibitem{Fol74} L. L. Foldy and R. A. Krajcik, {\it Phys. Rev. Lett.} {\bf 32},
1025, (1974); {\it Phys. Rev. D} {\bf 12}, 1700 (1975).
\bibitem{Gup65} V. K. Gupta,
B. S. Bhahar, and A. N. Mitra, {\it Phys. Rev. Lett.}
{\bf 15}, 974 (1965); V. S. Bhasin, H. Jacob, and A. N. Nitra, {\it Phys.
Lett.}
{\bf 32B}, 15 (1970); H. Jacob, V. S. Bhasin, and A. N. Mitra, {\it Phys. Rev.
D} {\bf 1}, 3496 (1970).
\bibitem{KLS89} L. A. Kondratyuk, F. M. Lev, and V. V. Soloview, {\it Few-Body
Systems} {\bf 7} 55 (1989).
\bibitem{Mul81} L. M\"{u}ller, {\it Nucl. Phys.} {\bf A360}, 331 (1981).
\bibitem{GLC86} W. Gl\"{o}ckle, T. S. H. Lee, and F. Coester, {\it Phys. Rev.
C}
{\bf 33}, 709 (1986).
\bibitem{Coe65} F. Coester, {\it Helv. Phys. Acta} {\bf 38}, 7 (1965).
\bibitem{SB51} E. E. Salpeter and H. A. Bethe,
 {\it Phys. Rev.} {\bf 84}, 1232 (1951).
\bibitem{AT66} A. Ahmadzadeh and J. A. Tjon,
{\it Phys. Rev.} {\bf 147}, 1111 (1966).
\bibitem{JT70} A. D. Jackson and J. A. Tjon, {\it Phys. Lett.} {\bf 32B},
9 (1970).
\bibitem{BS66} R. Blankenbecler and R. Sugar, {\it Phys. Rev.} {\bf 142}, 1051
(1966).
\bibitem{RT88} G. Rupp and J. A. Tjon, {\it Phys. Rev. C} {\bf 37}, 1729
(1988).
\bibitem{RT92} G. Rupp and J. A. Tjon, {\it Phys. Rev. C} {\bf 45}, 2133
(1992).
\bibitem{Mac89} R. Machleidt, {\it Adv. Nucl. Phys.} {\bf 19}, 189 (1989).
\bibitem{AO65} V. A. Alessandrini and R. L. Omnes,
 {\it Phys. Rev.}
{\bf 139}, B167 (1965).
\bibitem{FLN66} D. Z. Freedman, C. Lovelace, and J. M. Namyslowsky,
{\it Nuovo Cimento} {\bf 43A}, 258 (1966).
\bibitem{AAY68} R. Aaron, R. D. Amado, and J. E. Young,
 {\it Phys. Rev.} {\bf 174}, 2022 (1968).
\bibitem{RT77} A. S. Rinat and A. W. Thomas,
{\it Nucl. Phys.} {\bf A282}, 365 (1977).
\bibitem{detail} For more details, see Section 4.1 of Ref. [27].
\bibitem{foot3} For a definition of the three-body channels see, e.~g.,
Ref.~\cite{Bra88}.
\bibitem{foot1} The Bonn A and B potentials are presented in Section 4
and Appendix A, Table A.1, of Ref.~\cite{Mac89}.
\bibitem{Wglo77} W. Gl\"{o}ckle and R. Offermann, Phys. Rev. C {\bf 16},
2039 (1977); W. Gl\"{o}ckle, G. Hasberg, and A. R. Neghabian,
Z. Physik {\bf A305}, 217 (1982).
\bibitem{Bra88b} R. A. Brandenburg, G. S. Chulick, Y. E. Kim, D. J. Klepacki,
R. Machleidt, A. Picklesimer,
 and R. M. Thaler,
{\it Phys. Rev. C} {\bf 37}, 781
(1988).
\bibitem{foot2} See Fig.~4.4 of Ref.~\cite{Mac89}.
\bibitem{foot4} See Table 5.2 and Figs.~5.10--12 of Ref.~\cite{Mac89} where
phase parameters and observables as predicted
by the Paris  and the Bonn B potential are shown.
(Note that the Bonn B potential is denoted by `OBEP' in the cited Table and
Figures, while `Bonn' denotes the Bonn full model of Ref.~\cite{MHE87}.)
\bibitem{Mac92}
R. Machleidt, {\it XII. Int. Conf. on Few-Body Problems in Physics},
Vancouver, B. C., Canada, 1989, Contributed Papers, B. K. Jennings, ed.
(TRIUMF Report TRI-89-2) p. G51;
R. Machleidt {\it et al.}, submitted to {\it Few-Body Systems}.
\bibitem{Gro82} F. Gross, {\it Phys. Rev. C} {\bf 26}, 2203 (1982).
\end{thebibliography}
\end{document}